\documentclass[aps,prl,twocolumn,twoside,superscriptaddress,nofootinbib,preprintnumbers]{revtex4-1}

\usepackage{hyperref}
\usepackage{graphicx}
\usepackage{amsmath}
\usepackage{mathrsfs}
\usepackage{xspace}
\usepackage{color}

\newcommand{\df}{\mathrm{d}}

\def\cE{\mathcal{E}}

\newcommand{\si}{\sigma}

\def\cO{{\mathcal O}}

\newcommand{\nn}{\nonumber}

\allowdisplaybreaks[2]

\begin{document}


\title{Extending Precision Perturbative QCD with Track Functions}

\author{Yibei Li}
\email{yblee777@zju.edu.cn}
\affiliation{Zhejiang Institute of Modern Physics, Department of Physics, Zhejiang University, Hangzhou, 310027, China}

\author{Ian Moult}
\email{ian.moult@yale.edu}
\affiliation{Department of Physics, Yale University, New Haven, CT 06511}

\author{Solange Schrijnder van~Velzen}
\email{s.v.schrijndervanvelzen@uva.nl}
\affiliation{Nikhef, Theory Group,
	Science Park 105, 1098 XG, Amsterdam, The Netherlands}
\affiliation{Institute for Theoretical Physics Amsterdam and Delta Institute for Theoretical Physics, University of Amsterdam, Science Park 904, 1098 XH Amsterdam, The Netherlands}

\author{Wouter J.~Waalewijn}
\email{w.j.waalewijn@uva.nl}
\affiliation{Nikhef, Theory Group,
	Science Park 105, 1098 XG, Amsterdam, The Netherlands}
\affiliation{Institute for Theoretical Physics Amsterdam and Delta Institute for Theoretical Physics, University of Amsterdam, Science Park 904, 1098 XH Amsterdam, The Netherlands}

\author{Hua Xing Zhu}
\email{zhuhx@zju.edu.cn}
\affiliation{Zhejiang Institute of Modern Physics, Department of Physics, Zhejiang University, Hangzhou, 310027, China}

\begin{abstract}
Collider experiments often exploit information about the quantum numbers of final state hadrons to maximize their sensitivity, with applications ranging from the use of tracking information (electric charge) for precision jet substructure measurements, to flavor tagging for nucleon structure studies.
For such measurements, perturbative calculations in terms of quarks and gluons are insufficient, and nonperturbative track functions describing the energy fraction of a quark or gluon converted into a subset of hadrons (e.g., charged hadrons) must be incorporated.
Unlike fragmentation functions, track functions describe correlations between hadrons and therefore satisfy complicated nonlinear evolution equations whose structure has so far eluded calculation beyond the leading order.
In this \emph{Letter} we develop an understanding of track functions and their interplay with energy flow observables beyond the leading order,  allowing them to be used in state-of-the-art perturbative calculations for the first time. 
We identify a shift symmetry in the evolution of their moments that fixes their structure, and we explicitly compute the evolution of the first three moments at next-to-leading order, allowing for the description of up to three-point energy correlations.
We then calculate the two-point energy correlator on charged particles at $\cO(\alpha_s^2)$, illustrating explicitly that infrared singularities in perturbation theory are absorbed by moments of the track functions and also highlighting how these moments seamlessly interplay with modern techniques for perturbative calculations.
Our results extend the boundaries of traditional perturbative QCD, enabling precision perturbative predictions for energy flow observables sensitive to the quantum numbers of hadronic states. 
\end{abstract}

\maketitle

\emph{Introduction.}---The fundamental problem in collider experiments is understanding how the observed distribution of energy (energy flow) is carried by states of the underlying theory. For quantum chromodynamics (QCD), energy is carried by collimated sprays of hadrons, called jets. The energy flow within jets, referred to as jet substructure~\cite{Larkoski:2017jix,Asquith:2018igt,Marzani:2019hun}, has come to play a central role in modern collider experiments with wide-ranging applications from searches for new physics and studies of QCD in the vacuum at the LHC, to investigations of QCD in the medium produced by heavy-ion collisions, to future studies of nucleon structure at the electron-ion collider (EIC) \cite{Arratia:2019vju,Signori:2013mda,Li:2021txc}.

Understanding energy flow in confining theories such as QCD is particularly difficult because the microscopic degrees of freedom (the quarks and gluons, with which we can perform calculations using well-developed perturbative techniques) are different from the nonperturbative asymptotic states (baryons and mesons) observed in the detector.  In the absence of nonperturbative techniques for computing Lorentzian observables with real-time evolution, the traditional approach has been to focus on a restricted set of questions about the energy flow that can be computed in perturbation theory.  The famous theorems of Kinoshita \cite{Kinoshita:1962ur} and Lee and Nauenberg \cite{Lee:1964is} ensure that this is possible if one considers the energy flow summed over the quantum numbers of all possible final states. For such inclusive observables, there has been remarkable theoretical progress, driven both by advances in perturbative quantum field theory and by the development of more sophisticated techniques for the resummation of higher-order corrections in singular regions of phase space.

However, there are many interesting cases in modern collider experiments that require an understanding of energy flow on particular subsets of hadrons. For example, many state-of-the-art jet substructure measurements \cite{ATLAS:2015ytt,CMS:2018ypj,ATLAS:2019mgf,ATLAS:2020bbn,ALICE:2021njq}, as well as measurements of fragmentation \cite{ATLAS:2011myc,CMS:2014jjt,ALICE:2014dla,ATLAS:2017pgl,ALICE:2018ype,LHCb:2019qoc,ATLAS:2019dsv}, are performed on charged hadrons. This allows one to exploit the exceptional angular resolution of tracking detectors, as well as suppress pileup contamination, leading to more precise measurements. The ability to perform calculations on charged particles therefore increases the precision with which measurements can be performed in complicated hadronic environments. As another example, in DIS experiments such as HERA and the upcoming EIC, measurements of energy flow on flavored (e.g.,~strange) hadrons allow one to tag the initial state, providing new insight into nuclear structure \cite{Signori:2013mda,Li:2021txc}. New theoretical techniques are therefore required to go beyond the paradigm of fully inclusive energy flow observables.

In this \emph{Letter} we show that nonperturbative track functions \cite{Chang:2013rca,Chang:2013iba} can be used to extend recent progress in perturbative calculations to observables measured on subsets of final-state hadrons. We focus on a specific class of observables, the $N$-point energy correlators, which characterize correlations in energy flow. We will show that these observables are particularly advantageous for interfacing with track functions, as they only require the knowledge of the $\leq N$th moments of the track functions, instead of their full functional form. We show that the moments of the track function exhibit a shift symmetry, which significantly simplifies the structure of their evolution, and we present a method for computing their renormalization group equations (RGEs), along with explicit results for the first three moments at next-to-leading order. We then perform a complete calculation of the two-point energy correlator at $\mathcal{O}(\alpha_s^2)$, explicitly verifying that the infrared poles are as predicted by the track function evolution, and establishing for the first time the consistency of the track function formalism beyond leading order. This calculation also illustrates how moments of track functions interface with modern techniques for perturbative calculations. This opens the door to a wide variety of new calculations and enables powerful perturbative calculations to also be applied to obtain predications for certain nonperturbative measurements.

\emph{Energy correlation functions.}---Energy flow in collider experiments is characterized by correlation functions, $\langle \mathcal{E}(\vec n_1) \mathcal{E}(\vec n_2) \cdots \mathcal{E}(\vec n_k) \rangle$, of energy flow operators \cite{Sveshnikov:1995vi,Tkachov:1995kk,Korchemsky:1999kt,Bauer:2008dt,Hofman:2008ar,Belitsky:2013xxa,Belitsky:2013bja,Kravchuk:2018htv},
where the unit vectors $\vec n_i$ specify directions.
These correlation functions can be computed in perturbation theory and are known explicitly for the two- \cite{Basham:1978bw,Basham:1978zq,Belitsky:2013bja,Dixon:2018qgp,Luo:2019nig,Henn:2019gkr} and three-point correlators \cite{Chen:2019bpb}. They have recently received extensive theoretical interest from a variety of communities \cite{Hofman:2008ar,Belitsky:2013xxa,Belitsky:2013bja,Belitsky:2013ofa,Korchemsky:2015ssa,Belitsky:2014zha,Moult:2018jzp,Henn:2019gkr,Gao:2019ojf,Chicherin:2020azt,Kravchuk:2018htv,Kologlu:2019bco,Kologlu:2019mfz,1822249,Dixon:2019uzg,Moult:2019vou,Chicherin:2020azt,Chen:2020uvt,Chen:2020vvp,Chen:2019bpb,Chen:2020adz,Ebert:2020sfi,Chen:2021gdk,Korchemsky:2021okt}. In this \emph{Letter} we will emphasize another remarkable feature of these observables, namely, their simple interplay with nonperturbative track functions.

Assuming that the theory exhibits noninteracting asymptotic states, we can define a restricted energy flow operator $\mathcal{E}_R(\vec n)$ that only measures the energy flow associated with  states with certain quantum numbers. For example, $R$ can be the set of electrically charged hadrons. One can then consider the corresponding correlation function $\langle \mathcal{E}_R(\vec n_1) \mathcal{E}_R(\vec n_2) \cdots \mathcal{E}_R(\vec n_k) \rangle$. This correlation function cannot be computed purely in perturbation theory; however, it can be computed using moments of track functions,
\begin{align} \label{eq:ECF_tr}
 & \langle \mathcal{E}_R(\vec n_1) \mathcal{E}_R(\vec n_2) \cdots \mathcal{E}_R(\vec n_k) \rangle
  \\ & \quad
  = \sum\limits_{i_1, i_2,\cdots, i_k}\!\! T_{i_1}(1)\cdots T_{i_k}(1)  \langle \cE_{i_1} (\vec n_1) \cE_{i_2} (\vec n_2) \cdots \cE_{i_k} (\vec n_k) \rangle
\,.\nn\end{align}
Here $i_a = g, u, \bar u, d, \dots$ denote parton flavors and $T_{i_a}(1)$ is the first moment of the corresponding track function, discussed below. 
For simplicity we have not explicitly written contact terms. Their inclusion involves higher moments of the track functions, whose evolution is a focus of this \emph{Letter}.
This formula should be thought of as a timelike analog of the factorization into a partonic cross section and parton distribution functions (PDFs) for colliding protons. The precise definition of the operators $\cE_{i_1} (\vec n_1)$ can be given in terms of twist-2 quark and gluon operators of definite mass dimension (instead of the more familiar definite spin \cite{Hofman:2008ar}), as will be discussed elsewhere. For the purposes of this \emph{Letter} this matrix element can simply be thought of as the energy correlator computed for particular partonic states. This simple factorization formula for the energy correlators, which was first noted in \cite{Chen:2020vvp}, should be contrasted with other observables where a track function is required for each emitted quark or gluon \cite{Chang:2013rca,Chang:2013iba}, and the full functional form of the track function is required (one notable exception is~\cite{Chien:2020hzh}). While the recent resurgence of interest in the energy correlators was driven by their relation to correlation functions of local operators \cite{Hofman:2008ar,Belitsky:2013xxa,Belitsky:2013bja,Belitsky:2013ofa}, their relation to light-ray operators \cite{Hofman:2008ar,Kravchuk:2018htv,Kologlu:2019mfz,Kologlu:2019bco} and their simple perturbative structure \cite{Belitsky:2013ofa,Dixon:2018qgp,Henn:2019gkr}, for phenomenological applications at collider experiments it may indeed be Eq.~\eqref{eq:ECF_tr} that is their most important feature. This emphasizes the importance of the exchange of ideas between different communities to ultimately improve our understanding of real world QCD at colliders.

Beyond leading order in perturbation theory, the partonic energy correlation function on the right-hand side of Eq.~\eqref{eq:ECF_tr} contains infrared divergences. These are absorbed into the track functions, from which the track function renormalization group evolution (RGE) follows. Again, this is in exact analogy to the renormalization of the PDFs. The existence of a well-defined RGE is guaranteed by the universality of collinear limits, which can be proven to all orders using either diagrammatic techniques \cite{Kosower:1999xi,Feige:2014wja} or effective field theory \cite{Bauer:2001ct,Bauer:2000yr,Bauer:2001yt,Bauer:2002nz}. While PDFs and fragmentation functions have been well tested in higher-order calculations, track functions have only been investigated at leading logarithmic order. In order to firmly establish the track function formalism, we will compute their renormalization group evolution beyond the leading order, and furthermore show that these objects absorb the IR divergences appearing in an explicit perturbative calculation of the two-point correlator at $\mathcal{O}(\alpha_s^2)$.

\emph{Track functions and their symmetries.}--- Fragmentation functions \cite{Georgi:1977mg, Ellis:1978ty, Curci:1980uw,Collins:1981uk,Collins:1981uw}, which describe the energy distribution of \emph{single} hadrons, have a long history in QCD.
Track functions were originally introduced to describe the total momentum fraction of \emph{all} charged hadrons~\cite{Chang:2013rca,Chang:2013iba}, which is responsible for their more complicated evolution. They have been successfully applied in a number of calculations in perturbative QCD \cite{Chang:2013rca,Chang:2013iba,Chen:2020vvp,Chien:2020hzh}.  In this \emph{Letter} we generalize this notion to other subsets of hadrons specified by their quantum numbers, such that the track function for quarks in light-cone gauge is given by
\begin{align}\label{eq:track_def}
T_q(x)&=\int {\rm d}y^+ {\rm d}^2 y_\perp e^{ik^- y^+/2} \frac{1}{2N_c}\sum_{R, \bar R} \delta \left( x-\frac{P_R^-}{k^-}\right) \\
&\quad \times \text{tr} \left[  \frac{\gamma^-}{2} \langle 0| \psi(y^+,0, y_\perp)|R \bar R \rangle \langle R \bar R|\bar \psi(0) | 0 \rangle \right]\,,
\nn \end{align}
and similar for gluons.
Here $R$ denotes the hadrons in the final state  belonging to the subset, $\bar R$ denotes all other hadrons (the complement), and $P_R^-$ is the large light-cone momentum component of $R$. Despite the fact that we allow $R$ to be a more general subset of hadrons, we continue to refer to the object in Eq.~\eqref{eq:track_def} as a track function. Although similar to the definition of standard fragmentation, the fact that $R$ is the set of all hadrons of a given property leads to crucial differences.
Concretely, if $R$ consists of all pions in the final state, a final state with two pions with momentum fractions $x_1$ and $x_2$ would give a contribution $\sim \delta(x-x_1-x_2)$ to the track function, while it would give a contribution $\sim \delta(x-x_1) + \delta(x-x_2)$ to the pion fragmentation function.
Objects similar to track functions have been studied in the context of jet charge \cite{Waalewijn:2012sv,Krohn:2012fg} and fractal observables \cite{Elder:2017bkd}.

We will work with the $n$th moments of the track function, defined as $T_i(n,\mu)=\int\! \df x~ x^n~ T_i (x,\mu)$,
with the sum rule $T_i(0,\mu)=1$. In the following, we often suppress the argument $\mu$ for brevity. While the full track functions encode correlations between arbitrary numbers of hadrons, the $n$th moments can be thought of as encoding correlations between $n$ hadrons. More precisely, $T_i(n)$ is related to the $n$-hadron fragmentation function \cite{Konishi:1979cb,Sukhatme:1980vs,Majumder:2004wh} (this is discussed for $n=2$ in \cite{Waalewijn:2012sv}). 

\begin{figure}
  \centering
  \includegraphics[width=0.19\textwidth]{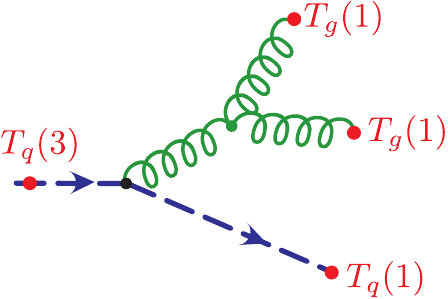} \qquad
  \includegraphics[width=0.19\textwidth]{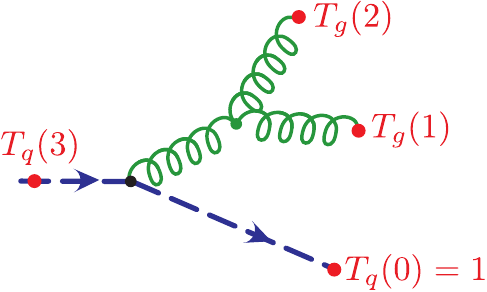}
  \caption{Triple collinear splittings contributing to the evolution of the track function moment  $T_q(3)$ at next-to-leading order: (a) $T_g(1) T_g(1)  T_q(1)$ and (b) $T_g(2) T_g(1)T_q(0)$. Here we have emphasized the appearance of $T_q(0)$, since it will play an important role in our understanding of the symmetries of the problem.}
  \label{fig:splittings}
\end{figure}

Because of the fact that they encode correlations between arbitrary numbers of hadrons, the track functions satisfy complicated nonlinear evolution equations, generated by multiparton splittings, see Fig.~\ref{fig:splittings}. Nothing is known about the structure of these equations beyond the leading order. To organize their structure, we note that unlike fragmentation functions which measure the energy fraction in a single hadron, track functions measure the energy fraction in all hadrons of a given type, as illustrated by the fact that there is a $T_i(m)$ on each branch of the splitting in Fig.~\ref{fig:splittings}. This implies that their evolution equations exhibit a shift symmetry, $T(x)\to T(x+a)$, corresponding to energy conservation. In moment space, this corresponds to an infinite set of polynomial shift symmetries, $T_i(1) \to T_i(1)-a$, $T_i(2) \to T_i(2)-2aT_i(1)+a^2 $, etc., which severely constrain the form of the evolution. 
In the body of this \emph{Letter} we focus on the  simplified case where the track functions are independent of quark flavor, and satisfy $T_{q}=T_{\bar q}$, to avoid the need for cumbersome notation. The most general case is presented in the \emph{Supplemental Material}. The shift symmetry implies that the evolution can be expressed in terms of shift-invariant \emph{central moments} $\sigma_i(2) = T_i(2) - T_i(1)^2$, $\sigma_i(3) = T_i(3) - 3T_i(2)T_i(1) +2 T_i(1)^3$, and so on, as well as $\Delta=T_q(1)-T_g(1)$. To simplify the notation, we will define $\vec \sigma(m)=(\sigma_q(m), \sigma_g(m))$. Using these symmetries, combined with a comparison to the fragmentation function limit, one can then show that
\begin{align}\label{eq:structure}
 \frac{\df}{\df\ln\mu^2}\Delta&= -(\gamma_{qq}(2)+\gamma_{gg}(2))  \Delta\,, \\
 \frac{\df}{\df\ln\mu^2} \vec \sigma(2) &=-\hat \gamma(3) \vec \sigma(2) + \vec \gamma_{\Delta^2}  \Delta^2\,, \nn\\
 \frac{\df}{\df\ln\mu^2} \vec \sigma(3) &=-\hat \gamma(4) \vec \sigma(3) + \hat \gamma_{\sigma_2 \Delta} \vec \sigma(2) \Delta+ \vec \gamma_{\Delta^3} \Delta^3\,, \nn\\
 \frac{\df}{\df\ln\mu^2} \vec \sigma(4) &=-\hat \gamma(5) \vec \sigma(4)+ \hat \gamma_{\sigma_2 \sigma_2}( \vec \sigma(2)\cdot\vec \sigma(2)^T)\nn \\
& \quad +\hat \gamma_{\sigma_3 \Delta} \vec \sigma(3) \Delta+ \hat \gamma_{\sigma_2 \Delta^2} \vec \sigma(2) \Delta^2+ \vec \gamma_{\Delta^4} \Delta^4\,, 
\nn\end{align}
and similar for the higher central moments. We emphasize that, since this structure is derived from symmetry, it holds to all orders in perturbation theory. It is remarkably simple compared to the most general form of the nonlinear evolution, emphasizing the important role the symmetry plays in constraining the evolution.
Here $\hat \gamma(J)$ are the matrices of timelike twist-2 spin-$J$ anomalous dimensions (Our conventions for these anomalous dimensions, as well as explicit results to next-to-leading order (NLO) are provided in the \emph{Supplemental Material}). The other anomalous dimensions are new, and we will explicitly compute them to next-to-leading order for the first three moments. Remarkably, despite the nonlinearity of the equations, shift invariance, combined with the uniqueness of the first three central moments, forces the evolution of the first three central moments of the track functions to be the standard DGLAP evolution~\cite{Gribov:1972ri,Gross:1974cs,Altarelli:1977zs,Dokshitzer:1977sg}, since (at least for charged hadrons) $\Delta=T_q(1)-T_g(1)$ is suppressed. On the other hand, unlike the central moments, the moments $T_i(n)$ exhibit complicated nonlinear evolution. The first genuine unsuppressed nonlinearities in the evolution of the central moments occur at the fourth moment due to mixings between $\vec \sigma(4)$ and $\vec \sigma(2)\cdot\vec \sigma(2)^T$, and similar at higher moments.

\emph{Next-to-leading order track function evolution.}---
We have computed the RGEs of the track functions by integrating the collinear splitting functions~\cite{Campbell:1997hg,Catani:1998nv,Kosower:2003np} to obtain a jet function for the $n$th moment of charged particles, differential in the invariant mass of all particles~\cite{Ritzmann:2014mka}. This is illustrated in Fig.~\ref{fig:splittings}. After the renormalization of this jet function (which is the same as the renormalization of the invariant mass jet function~\cite{Becher:2006qw,Becher:2010pd}), the track function evolution can be inferred from the remaining IR poles. Using the shift symmetry of the track function evolution equations significantly reduces the required calculations. However, as an additional check on our calculation, we have computed all mixing terms separately and verified that they respect the shift symmetry.

The complete results for the first three moments at NLO are given in the \emph{Supplemental Material}. Here we present only the mixing anomalous dimensions for the second moment to illustrate their features. We find
\begin{align}
    \gamma_{\Delta^2}^g&= \frac{4}{3}a_s n_f T_F \\
    &\quad + a_s^2 n_f T_F \left[   C_A\left(   -\frac{16}{9}\pi^2 +\frac{5897}{675} \right)+\frac{39161}{2700}C_F\right] \,, \nn\\
     \gamma_{\Delta^2}^q&=\frac{7}{6}a_s C_F+a_s^2 C_F\left[\frac{817}{108}C_A +\left( \frac{1369}{432}-\frac{14}{9}\pi^2 \right)C_F \right] \,,   \nn
\end{align}
where $a_s=\alpha_s/(4\pi)$. We have checked that the leading-order evolution of the second moment is consistent with that of the dihadron fragmentation function~\cite{Konishi:1979cb,Sukhatme:1980vs,deFlorian:2003cg}.

Because of the smallness of $\Delta$ for the specific case of charged hadrons in QCD, this implies that the NNLO corrections to the DGLAP anomalous dimensions are significantly larger than the NLO corrections to the mixing terms for the evolution of $\sigma(2)$. This allows one to immediately extend the evolution of $\sigma(2)$ to NNLO, using the known values of the timelike anomalous dimensions \cite{Chen:2020uvt,Mitov:2006ic,Moch:2007tx,Almasy:2011eq}. Combined with the  factorization formula of \cite{Dixon:2019uzg}, one can perform the resummation of the energy-energy correlator (EEC) on tracks in the collinear limit to next-to-next-to-leading logarithm, with eventual applications to jet substructure.

\emph{Two-point correlations at next-to-leading order.}---
Having calculated the evolution at next-to-leading order, we will now illustrate the consistency of track functions beyond the leading order. We will analytically compute the two-point energy correlation \cite{Basham:1978bw,Basham:1978zq,Richards:1982te} at  order $\mathcal{O}(\alpha_s^2)$, with a generic restriction on hadrons. The two-point energy correlator is characterized by a single angle $\chi$ between the two calorimeter cells. This calculation also illustrates how track functions seamlessly mesh with perturbative calculations, as this calculation matches the highest order available analytically for the two-point energy correlator on all partons.

The two-point energy correlator has been computed analytically at NLO for both $e^+e^-$ collisions \cite{Dixon:2018qgp} and Higgs decays \cite{Luo:2019nig,Gao:2020vyx}. When computed on all final-state particles, it is infrared finite to all orders in perturbation theory. However, when computed in dimensional regularization in $d=4-2\epsilon$, the partonic two-point correlator in Eq.~\eqref{eq:ECF_tr} has infrared poles in $\epsilon$ which must be absorbed into the track function. These poles are uniquely fixed in terms of the renormalization group (RG) evolution of the track function. Since the RG of the $n$th moments of the track function involve mixing with products of all lower moments, it is convenient to write their RG in an abstract form
\begin{align}
\frac{\df}{\df\ln \mu^2}\vec{ \bf T}_n=\widehat R_n~\vec {\bf T}_n\,,
\end{align}
where $\vec {\bf T}_n$ is a vector of all possible products of moments of track functions that have total weight $n$ (e.g.,~for $n=2$, $\vec {\bf T}_2=\{T_g(2),T_q(2), T_q(1)T_q(1), T_g(1)T_q(1),$ $T_g(1) T_g(1) \}$), and $\widehat R_n$ is a matrix, whose perturbative expansion is $\widehat R_n=\sum a_s^j \widehat R^{(j)}_n$. The IR divergences of the partonic energy correlators follow from the UV divergences of track functions
\begin{align}\label{eq:poles}
\vec {\bf T}_{n,\text{bare}}&=\vec {\bf{T}}_n(\mu)+a_s \frac{\widehat R^{(1)}_n}{\epsilon} \vec {\bf{T}}_n(\mu) \\
& \!+\!\frac{1}{2} a^2_s\left( \frac{\widehat R_n^{(2)}}{\epsilon}\!+\!\frac{ \widehat R_n^{(1)} \widehat R_n^{(1)}\!-\!\beta_0 \widehat R_n^{(1)}} {\epsilon^2}  \right) \vec {\bf{T}}_n(\mu)\!+\!\mathcal{O}(a_s^3).\nn
\end{align}
The $1/\epsilon^2$ poles at two loops are completely predicted from the one-loop renormalization, while the $1/\epsilon$ provide an independent calculation of the NLO RG evolution and the universality of the track functions.

To compute the EEC on tracks requires the calculation of the general partonic correlators in Eq.~\eqref{eq:ECF_tr}, extending the calculation of the EEC in \cite{Dixon:2018qgp}. To perform this calculation, we follow the approach of \cite{Dixon:2018qgp} and use reverse unitarity \cite{Anastasiou:2002yz} to express the phase space integrals for the EEC in terms of multiloop integrals. These integrals are reduced to master integrals using \textsc{LiteRed}~\cite{Lee:2012cn,Lee:2013mka} and \textsc{fire6}~\cite{Smirnov:2019qkx}. The master integrals are found to be the same as for the standard EEC and are evaluated by differential equations, using \textsc{canonica} \cite{Meyer:2017joq} to obtain their canonical form \cite{Henn:2013pwa}. Master integrals for the contact terms ($\delta(\chi)$) are the same as those for cut bubble integrals and can be extracted from \cite{Gehrmann-DeRidder:2003pne,Magerya:2019cvz}. The final results are written in terms of classical polylogarithms using \textsc{hpl}~\cite{Maitre:2005uu}, and complete analytic results will be presented elsewhere. We again emphasize that this perturbative calculation on tracks matches the state of the art for analytic perturbative calculations of energy flow observables. This clearly illustrates how important the factorization property of energy correlator observables in Eq.~\eqref{eq:ECF_tr} is for simplifying the perturbative component of calculations interfaced with tracks.

\begin{figure}
  \centering
    \includegraphics[width=0.49\textwidth]{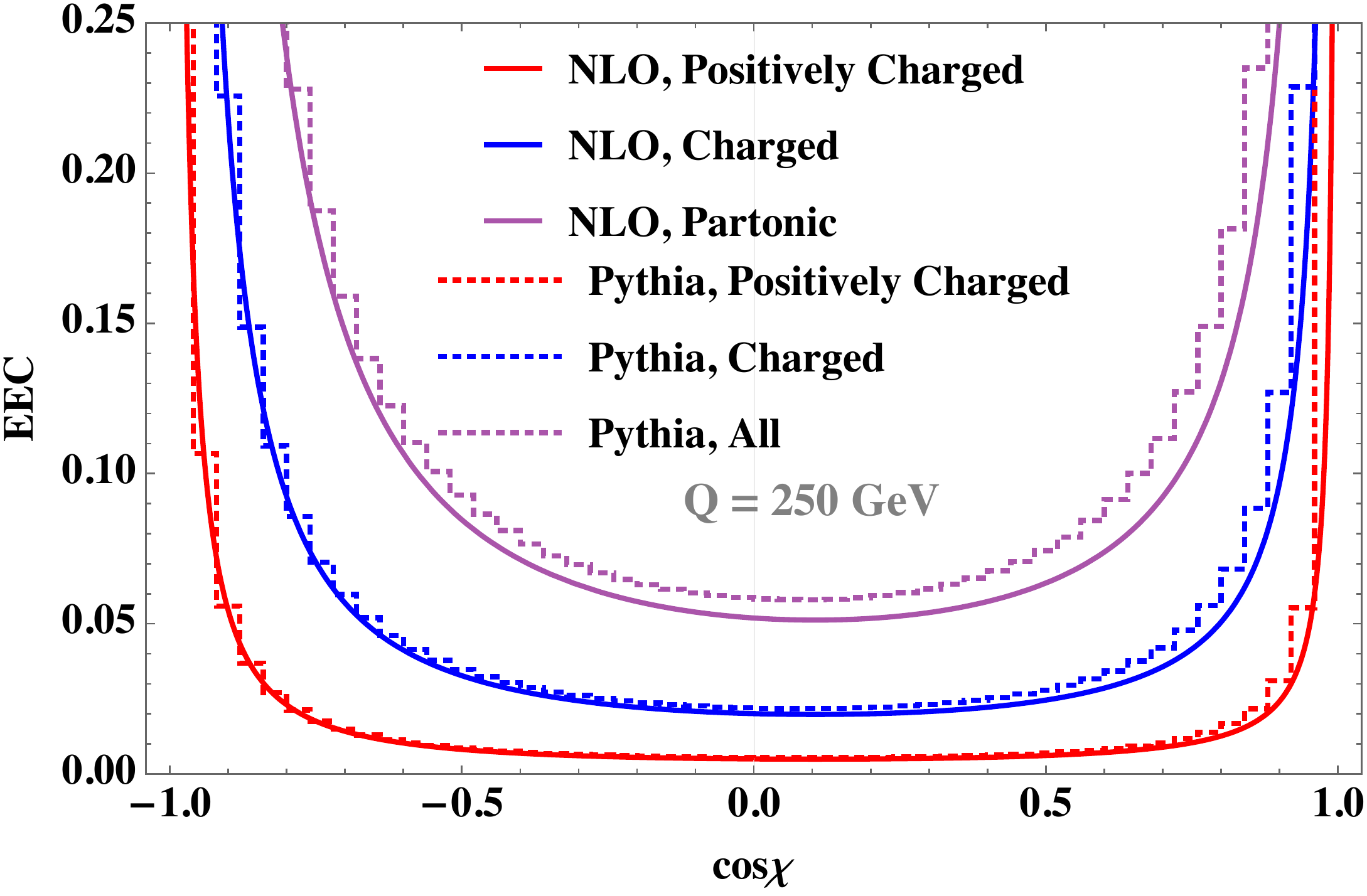}
  \caption{An illustration of the track function formalism for different subsets of final-state hadrons. Here we show the EEC at NLO as computed on all particles, on charged particles, and on positively charged particles, as well as a comparison with the Pythia parton shower. }
  \label{fig:EEC_all}
\end{figure}

\begin{figure}
  \centering
    \includegraphics[width=0.49\textwidth]{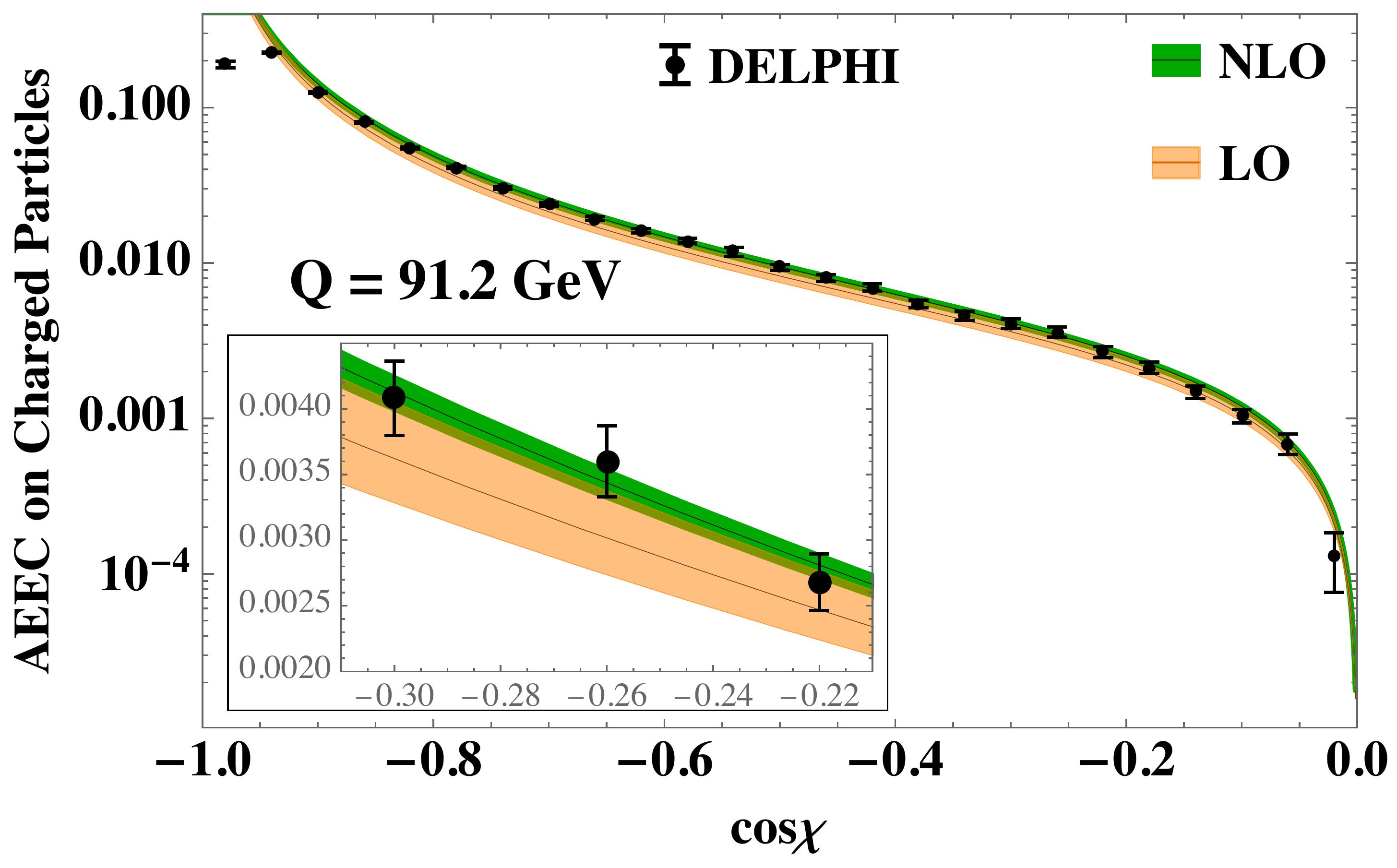}
  \caption{The AEEC on charged particles at LO and NLO, compared with DELPHI data. Excellent agreement is observed, except in the region $\cos(\chi)\to -1$, where resummation is required. The bands indicate the perturbative uncertainty from scale variations.}
  \label{fig:AEEC}
\end{figure}

Extracting the IR poles from the calculation of the partonic correlators, we find that they exactly match with those predicted by Eq.~\eqref{eq:poles}, providing a strong check on the track function formalism at  $\mathcal{O}(\alpha_s^2)$. Absorbing these poles into the renormalized track functions gives an IR finite result for the EEC computed on any subset of final-state hadrons at NLO. In Fig.~\ref{fig:EEC_all} we show our results for the EEC on all particles, charged particles, and positively charged particles, along with a comparison to Pythia~\cite{Sjostrand:2014zea}. Here and in Fig.~\ref{fig:AEEC} we have used track functions extracted from Pythia \cite{Chang:2013rca,Chang:2013iba}. In Fig.~\ref{fig:AEEC} we compare our LO and NLO results with DELPHI data \cite{DELPHI:1996sen} for the EEC asymmetry (AEEC), defined as $\text{AEEC}(\cos(\chi))=\text{EEC}(\cos(\chi))-\text{EEC}(-\cos(\chi))$, finding remarkably good agreement at NLO. The disagreement in the region $\cos(\chi) \to -1$ is due to the fact that we have not incorporated resummation. Such resummation could be included using the results of this \emph{Letter}, although it is beyond the scope of our current analysis. Although we find this agreement promising, there are a number of details about the normalization of the DELPHI data and the experimental analysis that must be better understood before more quantitative studies can be performed. This is the first $\cO(\alpha_s^2)$ calculation of a track-based observable, and we hope that the reduced experimental uncertainty for track-based observables can enable improved extractions of the strong coupling constant from event shapes.

\emph{Conclusions.}---In this \emph{Letter} we have extended the precision perturbative QCD program by showing the consistency of track functions beyond the leading order and elucidating aspects of their evolution and their interplay with energy flow observables. Our results allow one to harness the significant progress in perturbative quantum field theory to nonperturbative questions, allowing these to be computed beyond leading order for the first time, and avoiding the need to model these effects with parton showers. 

We believe that the results of this \emph{Letter} will have many applications to jet substructure, and QCD more generally, at a variety of colliders, ranging from the LHC, to ALICE, to EIC and $e^+e^-$ colliders, by drastically increasing the breadth of observables for which systematic perturbative calculations can be performed. We look forward to the phenomenological applications of our results.

\vspace{1em}
We thank Matt Leblanc, Ben Nachman, and Jennifer Roloff for many motivating discussions about the importance of tracks at the LHC, and Ben Nachman for help navigating the experimental literature.  We thank Hao Chen, Ming-xing Luo, Peter Jacobs, Petr Kravchuk, Tong Zhi Yang, Xiao Yuan Zhang, Jesse Thaler, Patrick Komiske, Meng Xiao, Jan Timmermans and Klaus Hamacher for helpful discussions. I.M. would like to thank both the KITP Santa Barbara and the Charles T. Munger Physics Residence for hospitality for the duration of this work.
Y.L. and H.X.Z. are supported by the National Natural Science Foundation of China under contract No. 11975200.
S.S.V.V. is supported by the NWO projectruimte 680-91-122.
W.W is supported by the ERC grant ERC-STG-2015-677323 and the D-ITP consortium, a program of NWO that is funded by the Dutch Ministry of Education, Culture and Science (OCW). 

\bibliography{spinning_gluon.bib}{}

\setcounter{equation}{0}
\renewcommand{\theequation}{S-\arabic{equation}}

\clearpage

\begin{widetext}

\section*{Supplemental Material}

In this \emph{Letter}, we presented the results for the NLO evolution of the first three moments of the track functions under the simplified assumption that $T_q=T_{\bar q}$, and that the track functions are equal for all quark flavors. This simplified case was sufficient to illustrate the structure of the equations, without excessive notation. In this \emph{Supplemental Material}, we present the complete results in several different forms. These different forms may be useful for different users, depending on their particular application. 

Although the results can be drastically simplified by working in terms of shift-invariant central moments, we begin by presenting the results for the evolution of the standard moments of the track functions. Using the following notation
\begin{align}
\frac{\df}{\df\ln \mu^2}{\color{blue} T_i(n)} = D_{T_i(n)}\,, \quad D_{T_i}(n) = 
\sum_{L=0}^\infty a_s^{L+1}D_{T_i(n)}^{(L)}\,,
\end{align}
we find that the evolution for the first three moments of the gluon track function is given by
\begin{align}\label{eq:full_gluon}
&D_{T_g(1)}^{(0)} =-\gamma^{(0)}_{gg}(2) {\color{blue}T_g{(1)}}-\sum_i\gamma^{(0)}_{qg}(2) {\color{blue}(T_{q_i}{(1)}+T_{\bar q_i}{(1)})}\,, \\
& D^{(0)}_{T_g(2)}= -\gamma^{(0)}_{gg}(3){\color{blue}T_g(2)}-\sum_{i}\gamma^{(0)}_{qg}(3){\color{blue}\left(T_{q_i}(2)+T_{\bar{q}_i}(2)\right)}+\frac{14}{5}C_A\,{\color{blue}T_{g}(1)T_{g}(1)}+\sum_{i}\frac{2}{5}T_F\,{\color{blue}T_{q_i}(1)T_{\bar{q}_i}(1)}
\,, \nn \\
& D^{(0)}_{T_g(3)}= -\gamma^{(0)}_{gg}(4){\color{blue}T_g(3)}-\sum_i \gamma^{(0)}_{qg}(4){\color{blue}\left(T_{q_i}(3)+T_{\bar{q}_i}(3)\right)}+\frac{21}{5}C_A\,{\color{blue}T_g(2)T_g(1)}+\sum_i\frac{3}{10}T_F {\color{blue}\left(T_{q_i}(2)T_{\bar{q}_i}(1)+T_{\bar{q}_i}(2)T_{q_i}(1)\right)}\,, \nn \\
&D_{T_g(1)}^{(1)} =-\gamma^{(1)}_{gg}(2) {\color{blue}T_g{(1)}}-\sum_i\gamma^{(1)}_{qg}(2) {\color{blue}(T_{q_i}{(1)}+T_{\bar q_i}{(1)})}\,, \nn\\
&D_{T_g(2)}^{(1)}=-\gamma^{(1)}_{gg}(3) {\color{blue}T_g{(2)}}-\sum_i\gamma^{(1)}_{qg}(3) {\color{blue}(T_{q_i}{(2)}+T_{\bar q_i}{(2)})}
+\left[  C_A^2 \left( -8 \zeta_3+\frac{26}{45}\pi^2 +\frac{2158}{675}  \right)-\frac{4}{9}C_A n_f T_F  \right] {\color{blue}T_g{(1)}T_g{(1)}} \nn  \\
&+\sum_i \left[ T_F\left( \!-\frac{299}{225}C_A\!-\!\frac{4387}{900}C_F   \right)  \right]  {\color{blue}T_g{(1)}(T_{q_i}{(1)} \!+\!T_{\bar q_i}{(1)})}
\!+\!\sum_i T_F\left[\left( \frac{12413}{1350}\!-\!\frac{52}{45}\pi^2  \right) C_A\!+\!\frac{1528}{225}C_F    \!-\!\frac{16}{25} n_f T_F \right]   {\color{blue}T_{q_i}{(1)} T_{\bar q_i}{(1)}} \,, \nn \\
&D_{T_g(3)}^{(1)} =-\gamma^{(1)}_{gg}(4) {\color{blue}T_g{(3)}}-\sum_i \gamma^{(1)}_{q g}(4) {\color{blue}(T_{q_i}{(3)}+T_{\bar q_i}{(3)})}+\left[ C_A^2 \left( 24 \zeta_3 -\frac{278}{15}\pi^2 +\frac{767263}{4500} \right)-\frac{2}{3}C_A n_f T_F   \right]{\color{blue}T_g(2)T_g{(1)}}   \nn \\
&+\sum_i \left[T_F \left( \!-\frac{46}{15}C_A\!-\!\frac{1727}{2250}C_F   \right)   \right] {\color{blue}T_g{(2)}(T_{q_i}{(1)}\!+\!T_{\bar q_i}{(1)})}
+ \sum_i T_F \left[\left( \frac{14}{15}\pi^2\!-\!\frac{10318}{1125}  \right)C_A\!-\!\frac{4544}{1125}C_F    \right]{\color{blue}(T_{q_i}{(2)}\!+\!T_{\bar q_i}{(2)}) T_g{(1)}}   \nn \\
&+\sum_i T_F\left[ \left( \frac{5321}{3000} -\frac{2}{5}\pi^2 \right)C_A+\frac{1523}{240}C_F-\frac{12}{25}n_f T_F  \right] {\color{blue}(T_{q_i}{(2)}T_{\bar q_i}{(1)}+T_{q_i}{(1)}T_{\bar q_i}{(2)} )}\nn \\
&+C_A^2 \left(-\frac{248561}{2250}+\frac{194}{15}\pi^2-24 \zeta_3  \right){\color{blue}T_g{(1)}T_g{(1)}T_g{(1)}} +\sum_i \left[ C_A T_f \left( \frac{23051}{1125}-\frac{28}{15}\pi^2   \right)-C_F T_f \frac{501}{100}   \right]{\color{blue}T_g{(1)}T_{q_i}{(1)} T_{\bar q_i}{(1)}}\,, \nn
\end{align}
while for quarks,
\begin{align}\label{eq:full_quark}
&D_{T_q(1)}^{(0)} =-\gamma^{(0)}_{gq}(2) {\color{blue}T_g(1)} -\gamma^{(0)}_{q q}(2) {\color{blue}T_q{(1)}}
\,,  \\
&D^{(0)}_{T_q(2)}=-\gamma_{gq}^{(0)}(3){\color{blue}T_g(2)}-\gamma^{(0)}_{qq}(3){\color{blue}T_q(2)}+3C_F\, {\color{blue}T_g(1)T_{q}(1)}
\,,\nn \\
&D^{(0)}_{T_q(3)}=-\gamma^{(0)}_{gq}(4){\color{blue}T_g(3)}-\gamma^{(0)}_{qq}(4){\color{blue}T_q(3)}+\frac{13}{10}C_F\,{\color{blue}T_g(2)T_q(1)}+\frac{16}{5}C_F\,{\color{blue}T_g(1)T_q(2)}
\,,\nn \\
&D_{T_q(1)}^{(1)} =-\gamma^{(1)}_{gq}(2) {\color{blue}T_g(1)} -\gamma^{(1)}_{q q}(2) {\color{blue}T_q{(1)}}-\gamma^{(1)}_{\bar q q}(2){\color{blue}T_{\bar q}{(1)} }-\sum_i \gamma^{(1)}_{Q q}(2){\color{blue}(T_{Q_i}{(1)}+ T_{\bar Q_i}{(1)})}
\,,\nn \\
&D_{T_q(2)}^{(1)}=-\gamma^{(1)}_{gq}(3) {\color{blue}T_g(2)} -\gamma^{(1)}_{q q}(3) {\color{blue}T_q{(2)}}-\gamma^{(1)}_{\bar q q}(3){\color{blue}T_{\bar q}{(2)}    }-\sum_i \gamma^{(1)}_{Q q}(3){\color{blue}(T_{Q_i}{(2)}+ T_{\bar Q_i}{(2)})} \nn \\
&+\left[ \left( \frac{1399}{5400}-\frac{7}{9}\pi^2  \right)C_A C_F -\frac{67}{18}C_F^2  \right]{\color{blue}T_g(1)T_g(1)}\nn \\
&+\left[ \left( -\frac{3023}{108}+\frac{34}{9}\pi^2-8\zeta_3   \right)C_A C_F +\left(\frac{3023}{54}-\frac{68}{9}\pi^2+16\zeta_3  \right) C_F^2 -\frac{53}{18}C_F T_F  \right] {\color{blue}T_q{(1)} T_q{(1)}}   \nn \\
&+\left[ \left(\frac{14057}{216}-\frac{77}{9}\pi^2+16 \zeta_3  \right)C_A C_F +\left(-\frac{14057}{108}+\frac{154}{9}\pi^2-32 \zeta_3  \right) C_F^2 -\frac{2803}{900}C_F T_F  \right] {\color{blue}T_q{(1)} T_{\bar q}{(1)}}    \nn \\
&+\left[  \frac{229}{18}C_A C_F+\left(\frac{2573}{72}-4\pi^2   \right)C_F^2 \right] {\color{blue}T_g(1) T_q{(1)}}
-\sum_i \frac{17}{100}C_F T_F {\color{blue}T_{Q_i}{(1)}T_{\bar Q_i}{(1)}}
-\sum_i  \frac{53}{18}C_F T_F {\color{blue}T_{q}{(1)}( T_{Q_i}{(1)}+ T_{\bar Q_i}{(1)}   )}\,, \nn \\
&D_{T_q(3)}^{(1)} =-\gamma^{(1)}_{gq}(4) {\color{blue}T_g(3)} -\gamma^{(1)}_{q q}(4) {\color{blue}T_q{(3)}}-\gamma^{(1)}_{\bar q q}(4){\color{blue}T_{\bar q}{(3)}   }-\sum_i \gamma^{(1)}_{Q q}(4){\color{blue}(T_{Q_i}{(3)}+ T_{\bar Q_i} {(3)}  )} \nn \\
&+\left[  -\frac{3787}{750} C_A C_F -\frac{249}{50}C_F^2  \right]{\color{blue}T_g(2)T_g(1)}   
+\left[  \left( \frac{7}{3}\pi^2-\frac{14161}{3000}  \right) C_A C_F +\left( \frac{84329}{6000}-\frac{26}{15}\pi^2 \right)C_F^2  \right] {\color{blue}T_g(2)T_q{(1)}}\nn \\
&+\left[ \frac{2327}{180}  C_A C_F +\left( \frac{10189}{250}-\frac{64}{15}\pi^2 \right)C_F^2  \right]{\color{blue}T_g(1)T_q{(2)}}    - \sum_i \frac{724}{225}C_F T_F  {\color{blue}T_q{(2)}( T_{Q_i}{(1)}+T_{\bar Q_i}{(1)})}\nn \\
&
-\sum_i \frac{9557}{9000}C_F T_F {\color{blue}T_q{(1)}( T_{Q_i}{(2)}+T_{\bar Q_i}{(2)})} - \sum_i \frac{59}{1000}C_F T_F {\color{blue}\left(T_{Q_i}{(2)} T_{\bar Q_i}{(1)} + T_{Q_i}{(1)} T_{\bar Q_i}{(2)}  \right)}     \nn \\
&+\left[ \left( -\frac{353801}{3600}+\frac{77}{6}\pi^2-24\zeta_3 \right)C_A C_F+ \left( \frac{353801}{1800}-\frac{77}{3}\pi^2+48\zeta_3 \right)C_F^2-\frac{12839}{3000}C_F T_F   \right] {\color{blue}T_q{(2)}T_q{(1)}}     \nn \\
&+\left[ \left( -\frac{369503}{3000}+\frac{77}{5}\pi^2-24\zeta_3 \right)C_A C_F+ \left(\frac{369503}{1500}-\frac{154}{5}\pi^2+48\zeta_3  \right)C_F^2-\frac{1261}{1125}C_F T_F   \right] {\color{blue}T_{\bar q}{(2)}T_q{(1)}}    \nn \\
&+ \left[ \left( \frac{649211}{6000}-\frac{139}{10}\pi^2+24\zeta_3 \right)C_A C_F+ \left(-\frac{649211}{3000}+\frac{139}{5}\pi^2-48 \zeta_3  \right)C_F^2-\frac{29491}{9000}C_F T_F   \right] {\color{blue}T_{\bar q}{(1)}T_q{(2)}}    \nn \\
&+\left[  \left( \frac{97883}{9000}-\frac{7}{3}\pi^2  \right)C_A C_F-\frac{181}{150}C_F^2  \right] {\color{blue}T_q{(1)}T_g(1)T_g(1)}
- \sum_i \frac{137}{500}C_F T_F {\color{blue}T_q{(1)}T_{Q_i}{(1)}T_{\bar Q_i}{(1)}}   \nn \\
&+\left[ \left( \frac{202651}{1800}-\frac{43}{3}\pi^2+24\zeta_3  \right)C_A C_F+\left( -\frac{202651}{900}+\frac{86}{3}\pi^2 -48\zeta_3  \right)C_F^2 -\frac{137}{500}C_F T_F  \right] {\color{blue}T_q{(1)}T_q{(1)}T_{\bar q}{(1)}}\,. \nn
\end{align}
Here $\gamma_{ij}^{(0)}(n)$ ($\gamma_{ij}^{(1)}(n)$) are the (N)LO moments of the timelike splitting function, and $Q\neq q$ is used to denote the distinct quark flavors.
The expressions for anti-quarks can be obtained by charge conjugation. 

As described in the \emph{Letter}, when written in terms of standard moments these equations are highly redundant, due to the presence of the underlying shift symmetry. To present the evolution equations in terms of central moments for the general case with different track functions for each quark flavor, we must extend the situation discussed in the \emph{Letter}, by introducing $\Delta_{q_i} = T_{q_i}(1) - T_g(1)$, in addition to $\sigma_j$ for each flavor.

For gluons, we find that the evolution of the second and third central moments, can be written as
\begin{align} \label{eq:final_gluon_shift}
  D^{(0)}_{\sigma_g(2)} & = -\gamma^{(0)}_{gg}(3){\color{blue}\sigma_g(2)}+\sum_i\left\{-\gamma^{(0)}_{qg}(3){\color{blue}\left(\sigma_{q_i}(2)+\sigma_{\bar{q}_i}(2)+\Delta_{q_i}^2+\Delta_{\bar{q}_i}^2\right)}
  + \frac{2}{5}T_F\, {\color{blue}\Delta_{q_i}\Delta_{\bar{q}_i}} \right\}\ , \\
  D^{(0)}_{\sigma_g(3)} &=  -\gamma^{(0)}_{gg}(4){\color{blue}\sigma_g(3)}+\sum_i\biggl\{-\gamma^{(0)}_{qg}(4){\color{blue}\left(\sigma_{q_i}(3)+\sigma_{\bar{q}_i}(3)
  +3\sigma_{q_i}(2)\Delta_{q_i}+3\sigma_{\bar{q}_i}(2)\Delta_{\bar{q}_i}
  +\Delta_{q_i}^3+\Delta_{\bar{q}_i}^3\right)} \nn\\ 
  &\quad  -2T_F\,{\color{blue}\sigma_g(2)(\Delta_{q_i}+\Delta_{\bar{q}_i})} 
  +\frac{3}{10}T_F\,{\color{blue}\left(\sigma_{q_i}(2)\Delta_{\bar{q}_i}+\sigma_{\bar{q}_i}(2)\Delta_{q_i}
  +\Delta_{q_i}^2\Delta_{\bar{q}_i}+\Delta_{\bar{q}_i}^2\Delta_{q_i} \right)}
  \biggr\}  \ ,\nn \\
D_{\si_g(2)}^{(1)} &=-\gamma^{(1)}_{gg}(3) {\color{blue}\si_g(2)}
+\sum_i \biggl\{-\gamma^{(1)}_{qg}(3) {\color{blue}(\si_{q_i}(2)+\si_{\bar q_i}(2)+\Delta_{q_i}^2+\Delta_{\bar q_i}^2)}
 \nn \\ 
& \quad
+T_F\Bigl[\Bigl( \frac{12413}{1350}-\frac{52}{45}\pi^2  \Bigr) C_A+\frac{1528}{225}C_F    -\frac{16}{25} n_f T_F \Bigr]   {\color{blue}\Delta_{q_i}\Delta_{\bar q_i}}
\biggr\} \,,\nn \\
D_{\si_g(3)}^{(1)} &=-\gamma^{(1)}_{gg}(4) {\color{blue}\si_g(3)}
+\sum_i \biggl\{-\gamma^{(1)}_{q g}(4) {\color{blue}(\si_{q_i}(3)+\si_{\bar q_i}(3) +3 \si_{q_i}(2) \Delta_{q_i}  +3\si_{\bar q_i}(2) \Delta_{\bar q_i} + \Delta_{q_i}^3 + \Delta_{\bar q_i}^3)}
\nn \\ & \quad
 +T_F \Bigl[\Bigl(- \frac{638}{45}+ \frac{8}{3}\pi^2\Bigr)C_A - \frac{3803}{250}C_F   \Bigr]  {\color{blue}\si_g(2)(\Delta_{q_i} + \Delta_{{\bar q}_i})}
\nn \\ &\quad 
+T_F\Bigl[ \Bigl( \frac{5321}{3000} -\frac{2}{5}\pi^2 \Bigr)C_A+\frac{1523}{240}C_F-\frac{12}{25}n_f T_F  \Bigr] 
{\color{blue}(\si_{q_i}(2) \Delta_{\bar q_i} +\si_{\bar q_i}(2) \Delta_{q_i}+ \Delta_{q_i}^2 \Delta_{\bar q_i} + \Delta_{\bar q_i}^2 \Delta_{q_i})}\biggr\}\,. \nn
\end{align}
This form emphasizes the large redundancy present in the expressions given in Eq.~\eqref{eq:full_gluon}.
We emphasize that while it is true that the mixing into $\sigma_{q_i}(2)$ and $\sigma_{q_i}(3)$ is governed to all loop order by $\gamma_{qg}$, the fact that the mixing into the products $\sigma_{q_i}(2) \Delta_{q_i}$ and $\Delta_{q_i}^3$ is also governed by this same anomalous dimension is a coincidence at this order in perturbation theory.

Finally, for the evolution of the quark track functions in terms of central moments, we have
\begin{align}\label{eq:final_quark_shift}
 D^{(0)}_{\sigma_q(2)} &=  -\gamma^{(0)}_{gq}(3){\color{blue}\left(\sigma_g(2)+\Delta_q^2\right)}-\gamma^{(0)}_{qq}(3){\color{blue}\sigma_q(2)} \,, \\
  D^{(0)}_{\sigma_q(3)} &=  -\gamma^{(0)}_{gq}(4){\color{blue}\left(\sigma_g(3)-3\sigma_g(2)\Delta_q-\Delta_q^3\right)}
  -\gamma^{(0)}_{qq}(4){\color{blue}\sigma_q(3)}+\frac{24}{5}C_F\, {\color{blue}\sigma_q(2)\Delta_q} \,, \nn \\
D_{\si_q(2)}^{(1)} & = - \gamma^{(1)}_{gq}(3)\, {\color{blue}\sigma_g(2)} - \gamma^{(1)}_{qq}(3) {\color{blue} (\sigma_{q}(2) + \Delta_{q}^2)}- \sum_{j} \gamma_{Qq}^{(1)} (3) {\color{blue} ( \sigma_{Q_j}(2) +\sigma_{\bar{Q}_j}(2) +\Delta_{Q_j}^2+ \Delta_{\bar{Q}_j}^2)}  \nn \\
&\quad - \gamma_{\bar{q}q}^{(1)} {\color{blue} (\sigma_{\bar{q}}(2)+ \Delta_{\bar{q}}^2 -2 \Delta_{q}\Delta_{\bar{q}})} +\frac{97}{54} C_F T_F \sum_{j} {\color{blue} \Delta_{q}(\Delta_{Q_j}+\Delta_{\bar{Q}_j})} 
 \nonumber \\
&\quad + \left[\frac{2957}{108} C_A C_F + \left(\frac{2323}{54}-\frac{64 \pi^2}{9}\right)C_F^2 + \left(\frac{97}{54}-\frac{256}{27} n_f \right) C_F T_F \right]  {\color{blue} \Delta_{q}^2} \nonumber \\
&\quad - \sum_{j} \frac{17}{100} C_F T_F{\color{blue} \Delta_{Q_j}\Delta_{\bar{Q}_j}}\,, \nn\\
D_{\si_q(3)}^{(1)} & = -\gamma_{gq}^{(1)}(4) {\color{blue}( \sigma_g(3)-2 \sigma_g(2)\Delta_{q})} -\gamma_{qq}^{(1)}(4){\color{blue} ( \sigma_{q}(3) - 2 \sigma_g(2) \Delta_{q} + 3\sigma_q(2) \Delta_{{q}}- 2 \Delta_{q}^3)} \nonumber \\
&\quad -\gamma_{\bar{q}q}^{(1)}(4) {\color{blue} (\sigma_{\bar{q}}(3)+\sigma_{g}(2) \Delta_{q} +3 \sigma_{\bar{q}}(2)\Delta_{\bar{q}} +3\sigma_{q}(2) \Delta_{\bar{q}} - 3\sigma_{\bar{q}}(2) \Delta_{q} +3\Delta_{q}^2 \Delta_{\bar{q}}} \nonumber \\ & \hspace{1.6cm} {\color{blue} -3 \Delta_{q}\Delta_{\bar{q}}^2+\Delta_{\bar{q}}^3 )} \nonumber \\
&\quad -\gamma_{Qq}^{(1)}(4) \sum_{j \neq i } {\color{blue}\bigl[ \sigma_{Q_j}(3)+ \sigma_{\bar{Q}_j}(3)-\sigma_g(2)\Delta_{q}+3\sigma_{Q_j}(2)\Delta_{Q_j}+3\sigma_{\bar{Q}_j}(2)\Delta_{\bar{Q}_j}} \nonumber \\ 
&\quad \hspace{2cm} {\color{blue}-3(\sigma_{Q_j}(2) +\sigma_{\bar{Q}_j}(2))\Delta_{q} + \Delta_{Q_j}^3+\Delta_{\bar{Q}_j}^3-3 \Delta_{q}( \Delta^2_{Q_j}+\Delta^2_{\bar{Q}_j}-\Delta_{Q_j}\Delta_{\bar{Q}_j})\bigr]
}\nonumber \\ 
&\quad-\frac{59}{1000} C_F T_F \sum_{j} {\color{blue}\bigl[\sigma_{Q_j}(2) \Delta_{\bar{Q}_j}+\sigma_{\bar{Q}_j}(2)\Delta_{Q_j} - (\sigma_{\bar{Q}_j}(2)+\sigma_{Q_j}(2) )\Delta_{q}+ \Delta_{Q_j}^2\Delta_{\bar{Q}_j} }\nonumber \\
&\quad \hspace{2.6cm} {\color{blue}+ \Delta_{Q_j}\Delta_{\bar{Q}_j}^2-\Delta_{q}(\Delta_{\bar{Q}_j}^2+\Delta_{Q_j}^2+\Delta_{Q_j}\Delta_{\bar{Q}_j}) \bigr] } \nonumber \\
&\quad + \frac{292}{75} C_F T_F\sum_{j} {\color{blue}\bigl[\sigma_{q}(2)(\Delta_{Q_j}+\Delta_{\bar{Q}_j})+\Delta_{q}^2(\Delta_{Q_j}+\Delta_{\bar{Q}_j})-\Delta_{q}\Delta_{Q_j}\Delta_{\bar{Q}_j} \bigr] }\nonumber \\
&\quad -\frac{97}{18}C_F T_F\sum_{j} {\color{blue}\bigl[\Delta_{q}^2(\Delta_{Q_j}+\Delta_{\bar{Q}_j})-\Delta_{q}\Delta_{Q_j}\Delta_{\bar{Q}_j} \bigr]}
- \frac{12929}{9000}(n_f-1) C_F T_F \, {\color{blue}\sigma_g(2) \Delta_{q}}\nonumber \\
&\quad+ \left[ \frac{29}{300}C_A C_F-\frac{29}{150}C_F^2+\frac{5797}{1125}C_F T_F   \right]{\color{blue}\sigma_{q}(2) \Delta_{\bar{q}} }\nonumber \\ 
&\quad + \left[\left(  -\frac{12929}{9000}C_F +\frac{4648}{225}C_F n_f \right)T_F +\left( -\frac{2163833}{18000}+\frac{247}{30}\pi^2-12\zeta_3 \right) C_A C_F \right.\nn \\
&\quad \hspace{2cm}\left. +\left(\frac{81443}{3000}-\frac{23}{15}\pi^2+24\zeta_3  \right) C_F^2  \right] {\color{blue} ( \sigma_{g}(2)\Delta_{q}+\Delta_{q}^3)}\nonumber \\
&\quad + \left[   \frac{45253}{450}C_A C_F+C_F^2\left( \frac{662327}{3600}-\frac{82}{3}\pi^2 \right)+\left(  \frac{23719}{4500}C_F-\frac{671}{18}C_F n_f \right) T_F \right]{\color{blue}\sigma_q(2) \Delta_q}\,. \nn
\end{align}
This case is notationally more cumbersome than for the gluon evolution due to the contributions from different quark flavors. As with Eq.~\eqref{eq:final_gluon_shift}, this result exhibits a number of coincidences in the evolution, that will not persist at higher orders in perturbation theory.

For completeness, we also provide results for the timelike anomalous dimensions appearing in the evolution equations for the track functions. Expanding the timelike splitting functions perturbatively in $a_s=\alpha_s/(4\pi)$ as
\begin{align}
P_{ij}(z)=\sum_{L=0}^\infty a_s^{L+1}P_{ij}^{(L)}(z)\,,
\end{align}
we define the Mellin moments of the timelike splitting functions as
\begin{align}
\gamma_{ij}^{(L)}(k)=-\int_0^1 \df z\ z^{k-1}P_{ij}^{(L)}(z)\,.
\end{align}
This definition is chosen so that for the case of the spacelike splitting function, one obtains the standard twist-2 spin-$k$ anomalous dimensions. We obtained our results by directly integrating the $z$-space results of \cite{Chen:2020uvt}. This has the advantage that it works for both even and odd $k$.

At LO we have,
\begin{align}
\gamma_{gg}^{(0)}(2)&=\frac{4}{3}n_f T_F\,,\quad\gamma_{gg}^{(0)}(3)=\frac{14}{5}C_A+\frac{4}{3}n_fT_F\,, \quad \gamma_{gg}^{(0)}(4)=\frac{21}{5}C_A+\frac{4}{3}n_fT_F\,, \nonumber\\
\gamma_{gq}^{(0)}(2)&=-\frac{8}{3}C_F\,,\quad\gamma_{gq}^{(0)}(3)=-\frac{7}{6}C_F\,, \quad \gamma_{gq}^{(0)}(4)=-\frac{11}{15}C_F\,,\nonumber\\
\gamma_{qg}^{(0)}(2)&=-\frac{2}{3}T_F\,,\quad\gamma_{qg}^{(0)}(3)=-\frac{7}{15}T_F\,,\quad \gamma_{qg}^{(0)}(4)=-\frac{11}{30}T_F\,,  \nonumber\\
\gamma_{qq}^{(0)}(2)&=\frac{8}{3}C_F\,,\quad\gamma_{qq}^{(0)}(3)=\frac{25}{6}C_F\,, \quad \gamma_{qq}^{(0)}(4)=\frac{157}{30}C_F\,,\nonumber\\
\gamma_{\bar{q}q}^{(0)}(2)&=\gamma_{\bar{q}q}^{(0)}(3)=\gamma_{\bar{q}q}^{(0)}(4)=0\,,\nonumber\\
\gamma_{Qq}^{(0)}(2)&=\gamma_{Qq}^{(0)}(3)=\gamma_{Qq}^{(0)}(4)=0\,,\nn \\
\gamma_{\bar{Q}q}^{(0)}(2)&=\gamma_{\bar{Q}q}^{(0)}(3)=\gamma_{\bar{Q}q}^{(0)}(4)=0\,.
\end{align}
At NLO we have, 
\begin{align}
\gamma_{gg}^{(1)}(2)&=n_f T_F \left[\left(\frac{200}{27}-\frac{16 \pi ^2}{9}\right)C_A+\frac{260}{27}C_F\right]\,,\nonumber\\
\gamma_{gq}^{(1)}(2)&=\left(\frac{32 \pi ^2}{9}-\frac{568}{27}\right)C_F^2-\frac{376}{27}C_AC_F \,,\nonumber\\
\gamma_{qg}^{(1)}(2)&=T_F \left[\left(\frac{8 \pi ^2}{9}-\frac{100}{27}\right)C_A-\frac{130}{27}C_F\right]\,,\nonumber\\
\gamma_{qq}^{(1)}(2)&=C_AC_F \left(4 \zeta_3+\frac{1495}{54}-\frac{17 \pi^2}{9}\right)+C_F^2 \left(-8 \zeta_3-\frac{175}{27}+\frac{2 \pi^2}{9}\right)-\frac{128}{27}C_Fn_fT_F+\frac{64}{27}C_FT_F\,,\nonumber\\
\gamma_{\bar{q}q}^{(1)}(2)&=C_AC_F \left(-4 \zeta_3-\frac{743}{54}+\frac{17 \pi^2}{9}\right)+C_F^2 \left(8 \zeta_3+\frac{743}{27}-\frac{34\pi^2}{9}\right)+\frac{64}{27}C_FT_F \,,\nonumber\\
\gamma_{Qq}^{(1)}(2)&=\frac{64}{27}C_FT_F\,,\nonumber\\
\gamma_{\bar{Q}q}^{(1)}(2)&=\frac{64}{27}C_FT_F \,,\nonumber\\
\gamma_{gg}^{(1)}(3)&=C_A^2 \left(-8 \zeta_3+\frac{2158}{675}+\frac{26 \pi ^2}{45}\right)+n_fT_F \left[\left(\frac{3803}{675}-\frac{16 \pi ^2}{9}\right)C_A+\frac{12839 }{2700}C_F\right] \,,\nonumber\\
\gamma_{gq}^{(1)}(3)&=\left(-\frac{39451}{5400}-\frac{7 \pi ^2}{9}\right)C_AC_F+\left(\frac{14\pi ^2}{9}-\frac{2977}{432}\right) C_F^2 \,,\nonumber\\
\gamma_{qg}^{(1)}(3)&=T_F \left[\left(\frac{619}{2700}+\frac{14 \pi ^2}{45}\right) C_A-\frac{833}{216}C_F\right]-\frac{8}{25}n_fT_F^2 \,,\nonumber\\
\gamma_{qq}^{(1)}(3)&=C_AC_F \left(4 \zeta_3+\frac{16673}{432}-\frac{43 \pi^2}{18}\right)+C_F^2 \left(-8 \zeta_3+\frac{989}{432}-\frac{7 \pi^2}{9}\right)-\frac{415}{54}C_Fn_fT_F+\frac{4391}{5400}C_FT_F\,,\nonumber\\
\gamma_{\bar{q}q}^{(1)}(3)&=C_AC_F \left(4 \zeta_3+\frac{8113}{432}-\frac{43 \pi^2}{18}\right)+C_F^2 \left(-8 \zeta_3-\frac{8113}{216}+\frac{43 \pi^2}{9}\right)+\frac{4391}{5400}C_FT_F \,,\nonumber\\
\gamma_{Qq}^{(1)}(3)&=\frac{4391}{5400}C_FT_F\,,\nonumber\\
\gamma_{\bar{Q}q}^{(1)}(3)&=\frac{4391}{5400}C_FT_F \,, \nn \\
\gamma_{gg}^{(1)}(4)&=\left(\frac{90047}{1500}-\frac{28 \pi ^2}{5}\right)C_A^2+n_fT_F\left[\left(\frac{2273}{675}-\frac{16 \pi ^2}{9}\right)C_A+\frac{57287}{13500}C_F\right]\,,\nonumber\\
\gamma_{qg}^{(1)}(4)&=T_F\left[\left(\frac{22 \pi ^2}{45}-\frac{60391}{27000}\right)C_A-\frac{166729}{54000}C_F\right]-\frac{12}{25}n_fT_F^2\,,\nonumber\\
\gamma_{gq}^{(1)}(4)&=\left(\frac{44 \pi ^2}{45}-\frac{104389}{27000}\right) C_F^2-\frac{142591}{13500}C_AC_F\,,\nonumber\\
\gamma_{qq}^{(1)}(4)&=C_AC_F\left(4 \zeta_3+\frac{2495453}{54000}-\frac{247 \pi^2}{90}\right)+C_F^2\left(-8 \zeta_3+\frac{55553}{6000}-\frac{67 \pi^2}{45}\right)-\frac{13271}{1350}C_Fn_fT_F\nonumber\\
&+\frac{11867}{27000}C_FT_F\,,\nonumber\\
\gamma_{\bar{q}q}^{(1)}(4)&=C_AC_F\left(-4 \zeta_3-\frac{1202893}{54000}+\frac{247 \pi^2}{90}\right)+C_F^2\left(8 \zeta_3+\frac{1202893}{27000}-\frac{247 \pi^2}{45}\right)+\frac{11867}{27000}C_FT_F\,,\nonumber\\
\gamma_{Qq}^{(1)}(4)&=\frac{11867}{27000}C_FT_F\,,\nonumber\\
\gamma_{\bar{Q}q}^{(1)}(4)&=\frac{11867}{27000}C_FT_F\,.
\end{align}

\end{widetext}

\end{document}